\documentclass[aps,prl,twocolumn,10pt,showpacs,superscriptaddress,amsmath,amssymb,nobibnotes]{revtex4-1}

\usepackage{graphicx}
\usepackage{amsfonts}    % need for subequations
\usepackage{graphicx}   % need for figures
\usepackage{verbatim}   % useful for program listings
\usepackage{color}      % use if color is used in text
\usepackage{subfigure}  % use for side-by-side figures
\usepackage{hyperref}   % use for hypertext links, including those to external documents and URLs
\usepackage{natbib}
\usepackage{booktabs}
\usepackage{threeparttable}
\usepackage{dcolumn}
\usepackage{multirow}
\usepackage{float}

\begin{document}
\title{Experimental Realization of Sequential Weak Measurements of Arbitrary Non-commuting Pauli Observables}
\author{Jiang-Shan Chen}
\author{Meng-Jun Hu}\email{mengjun@mail.ustc.edu.cn}
\author{Xiao-Min Hu}
\author{Bi-Heng Liu}\email{bhliu@ustc.edu.cn}
\author{Yun-Feng Huang}
\author{Chuan-Feng Li}
\author{Guang-Can Guo}
\author{Yong-Sheng Zhang}\email{yshzhang@ustc.edu.cn}
\affiliation{CAS Key Laboratory of Quantum Information, University of Science and Technology of China, Hefei, 230026, People's Republic of China}
\affiliation{Synergetic Innovation Center of Quantum Information and Quantum Physics, University of Science and Technology of China, Hefei, 230026, People's Republic of China}

\date{\today}

\begin{abstract}
Sequential weak measurements of non-commuting observables is not only fundamentally interesting in quantum measurement but also shown potential in various applications. The previous reported methods, however, can only realize limited sequential weak measurements experimentally. In this Letter,  we propose the realization of sequential measurements of arbitrary observables and experimentally demonstrate for the first time the measurement of sequential weak values of three non-commuting Pauli observables by using genuine single photons. The results presented here will not only improve our understanding of quantum measurement, e.g. testing quantum contextuality, macroscopic realism, and uncertainty relation, but also have many applications such as realizing counterfactual computation, direct process tomography, direct measurement of the density matrix and unbounded randomness certification.  
\end{abstract}

\maketitle
{\it Introduction.}
Limited by Heisenberg uncertainty principle \cite{Heisenberg}, conventional quantum measurement theory forbids joint information extraction for non-commuting observables since that measurement inevitably collapses system into one of eigenstates of observable  . This awkward situation can be partially relaxed, however, by weak measurements in which lesser information is obtained but with smaller disturbance to system.  Weak measurement, which came up by Aharonov, Albert and Vaidman (AAV) in their 1988 original paper \cite{1}, has attracted widely interests in recent years. When the interaction between quantum system and probe is weak enough and a proper post-selected state is given to quantum system after interaction, the AAV shows that there appears a strange value of observable which they call it weak value (WV). Specifically, the WV of observable $\hat{A}$ is defined as $\langle\hat{A}\rangle_{w}={\langle\psi_{f}|\hat{A}|\psi_{i}\rangle}/{\langle\psi_{f}|\psi_{i}\rangle}$ with $|\psi_{i}\rangle$ and $|\psi_{f}\rangle$ represent pre-selected and post-selected state of quantum system respectively. The WV is full of strangeness even from the point view of standard quantum theory. It is in general a complex value and can beyond the eigenvalue spectrum of observable. Despite of the long-standing controversy about the physical meaning of WV \cite{2,3,4}, the strange properties of WV itself show its powerful applications in solving quantum paradox \cite{5}, quantum state reconstruction \cite{6,7,Lundeen} and signal amplification \cite{8,9,10}.

The fact that one hardly disturb the system when making weak measurements implies that non-commuting observables can be measured in succession and joint properties can be extracted via so called sequential weak values (SWVs) \cite{sequential, weak}. The measurement of SWVs experimentally is thus of great interest from both fundamental and applicative point of view. Unfortunately, most experiments reported till now have only realized single observable weak measurements or joint weak measurements on commuting observables \cite{single1, single2, single3, single4,single5} and the first experimental measurement of SWVs of two incompatible observables in photonic system is reported only very recently \cite{weak, pro2}, in which case the use of spatial distribution of photons as the pointer restrict its ability to perform further sequential measurements. The realization of measurement of SWVs, both real part and imaginary part, of arbitrary non-commuting observables is still missing. 

In this Letter, we propose the realization of measurement of SWVs of arbitrary non-commuting observables and experimentally demonstrate the first measurement of SWVs of three non-commuting Pauli observables in photonic system with genuine single photons. The core of our proposal is the realization of sequential weak measurements (SWMs) of arbitrary observables. Contrary to previous methods, the use of discrete pointer in our case makes the proposal here more feasible practically.
The results presented here are not only helpful in understanding measurement of non-commuting observables but also useful in realizing counterfactual computation experimentally \cite{sequential, coun1, coun2}, direct process tomography \cite{pro1, pro2}, direct measurement of density matrix \cite{Lundeen} and unbounded randomness certification \cite{ran1, ran2}. 
%Recently, the Leggett-Garg inequality (LGI) \cite{LG}, which served as test for the predictions of quantum mechanics for macroscopic variables, has been generalized to weak measurements \cite{G1, G2, G3} and the violation of the generalized LGI has been reported \cite{single5, V1}. Using SWMs here, we also observe the violation of Leggett-Garg inequality by up to $10$ standard deviations. 

\begin{figure*}[tbp]
\centering
\includegraphics[scale=0.4]{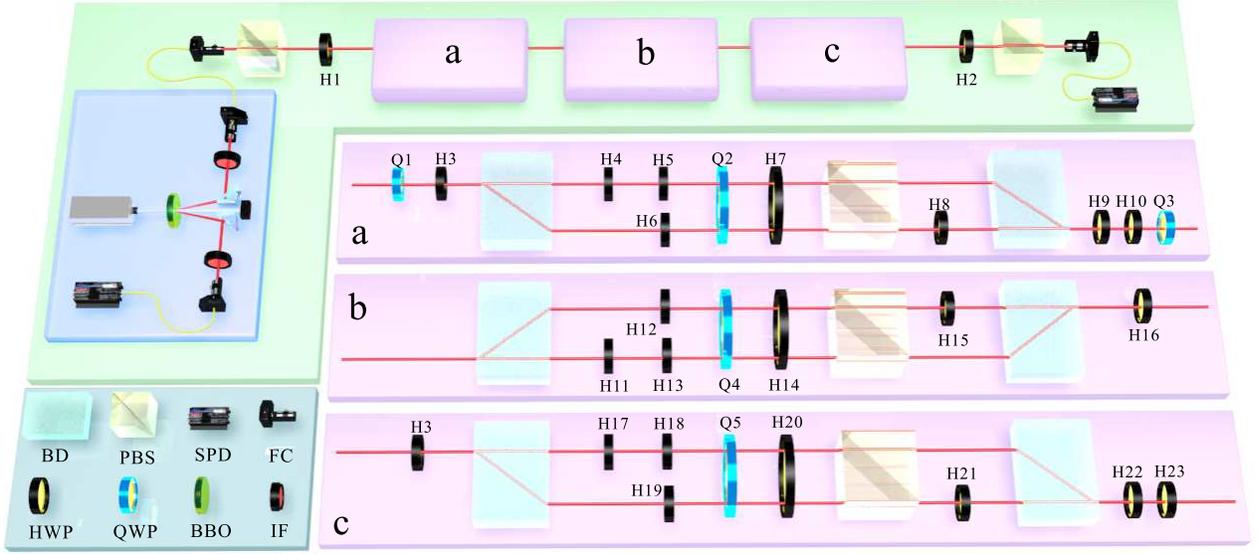}
\caption{Experimental setup for realizing sequential weak measurements of three non-commuting polarization observables of photons. The single photons is produced by generating pair of photons via SPDC process with idler photons used as trigger. The signal photons, after initial state preparation, is sent into the weak measurement module a, b, and c sequentially. After sequential weak measurements, post-selection is performed and coincidence counting is used. The module a, b and c realize weak measurements of polarization observables $\hat{\sigma}_{y}, \hat{\sigma}_{z}$ and $\hat{\sigma}_{\varphi}$ respectively. }
\label{fig.1}
\end{figure*}

{\it Sequential weak measurements and sequential weak values.}
We begin with a brief review of SWMs and SWVs. In a typical weak measurement, the system with pre-selected state is first weakly coupled to a pointer and then post-selected into a specific state. The real part and imaginary part of weak value of measured observable can be obtained respectively when we perform measurement of conjugate observables of pointer e.g., position or momentum. In most theoretical discussions and experiments, pointers are chosen as Gaussian continuous distribution with the von Neumann-type interaction Hamiltonian $\hat{H}=g\hat{A}\otimes\hat{P}$ \cite{1,Lundeen,weak}.  
Here we adopt discrete pointer e.g., polarization degree of freedom of photons and the interaction Hamiltonian has the form $\hat{H}=\gamma\hat{A}\otimes\hat{\sigma}_{y}$, in which eigenstates of observable $\hat{A}$ cause different rotations of pointer \cite{6,strong}. In the case of continuous pointer, the number of SWMs is limited due to the fact that only three independent spatial degrees of freedom are available \cite{weak}, while it is not a problem in the case of discrete pointer as we will show in the experiment part.
The state of pointer, after system is post-selected into state $|\psi_{f}\rangle$, becomes (unnormalized)
\begin{equation}
|\tilde{\varphi}_{p}\rangle=\langle\psi_{f}|e^{-i\gamma\hat{A}\otimes\hat{\sigma}_{y}}|\psi_{i}\rangle|0_{p}\rangle,
\end{equation}
where $|0_{p}\rangle$ is the initial state of pointer and natural unit is used that $\hbar\equiv1$. The pointer belongs to a qubit space spanned by the orthogonal states $\lbrace|0_{p}\rangle,|1_{p}\rangle\rbrace$. When the coupling $\gamma$ is weak enough the pointer state $|\tilde{\varphi}_{p}\rangle$ can be approximately rewritten as
\begin{equation}
|\tilde{\varphi}_{p}\rangle\approx\langle\psi_{f}|\psi_{i}\rangle e^{-i\gamma\langle\hat{A}\rangle_{w}\otimes\hat{\sigma}_{y}}|0_{p}\rangle
\end{equation}
with $\langle\hat{A}\rangle_{w}={\langle\psi_{f}|\hat{A}|\psi_{i}\rangle}/{\langle\psi_{f}|\psi_{i}\rangle}$ is weak value.
The post-selection of system causes $\theta\langle\hat{A}\rangle_{w}$ rotation of pointer in the weak coupling case and $\langle\hat{A}\rangle_{w}$ can be extracted by performing measurement of conjugate observables of pointer. The real part and imaginary part of $\langle\hat{A}\rangle_{w}$ are obtained from the expectation value of pointer observables $\hat{\sigma}_{+}$ and $\hat{\sigma}_{R}$ respectively. Specifically, we have
\begin{equation}
\begin{split}
\langle\hat{\sigma}_{+}\rangle_{p}&=2\gamma\mathrm{Re}\langle\hat{A}\rangle_{w} \\
\langle\hat{\sigma}_{R}\rangle_{p}&=2\gamma\mathrm{Im}\langle\hat{A}\rangle_{w},
\end{split}
\end{equation}
where $\hat{\sigma}_{+}\equiv|+\rangle\langle+|-|-\rangle\langle -|$, $\hat{\sigma}_{R}\equiv|R\rangle\langle R|-|L\rangle\langle L|$ and $|\pm\rangle=(|0\rangle\pm|1\rangle)/\sqrt{2}$, $|R\rangle=(|0\rangle+i|1\rangle)/\sqrt{2}, |L\rangle=(|0\rangle-i|1\rangle)/\sqrt{2}$. 

\begin{figure*}[tbp]
\centering
\subfigure[]{\label{Fig.2.subf.1}
\includegraphics[scale=0.468]{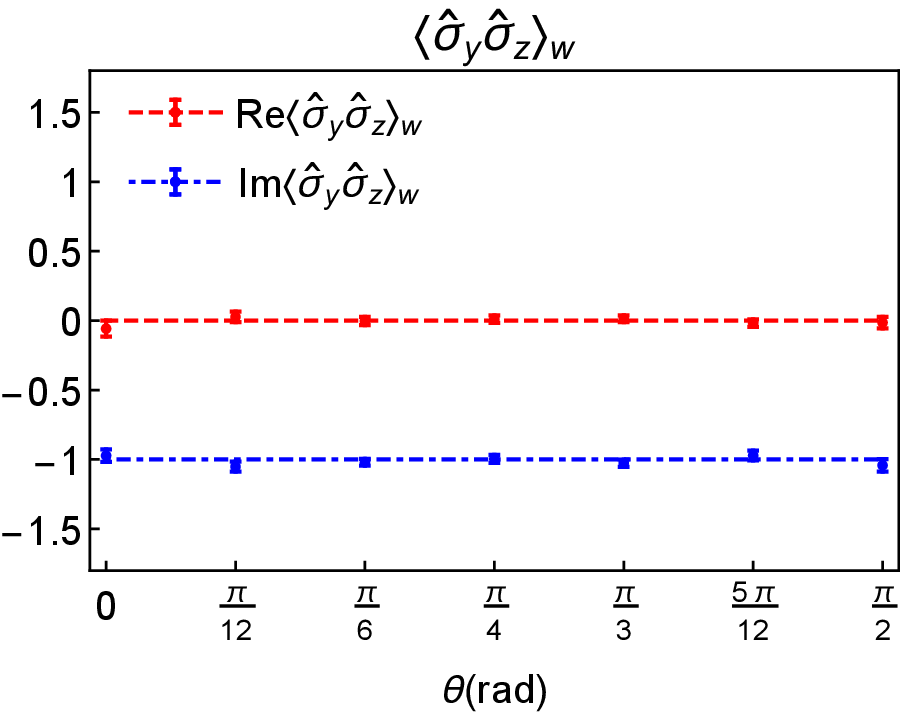}}
\subfigure[]{\label{Fig.2.subf.2}
\includegraphics[scale=0.468]{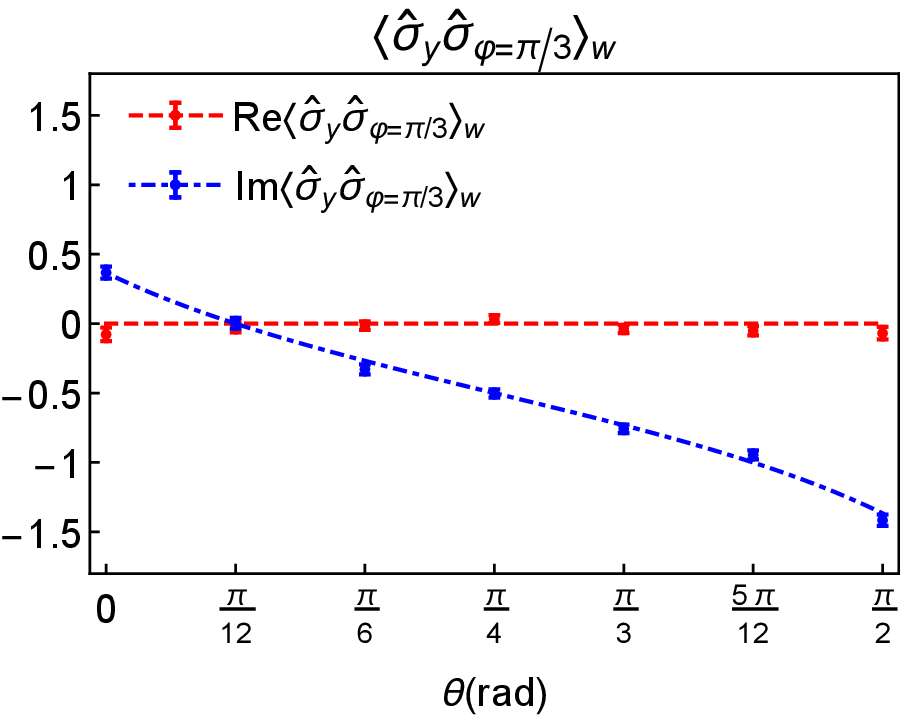}}
\subfigure[]{\label{Fig.2.subf.2}
\includegraphics[scale=0.468]{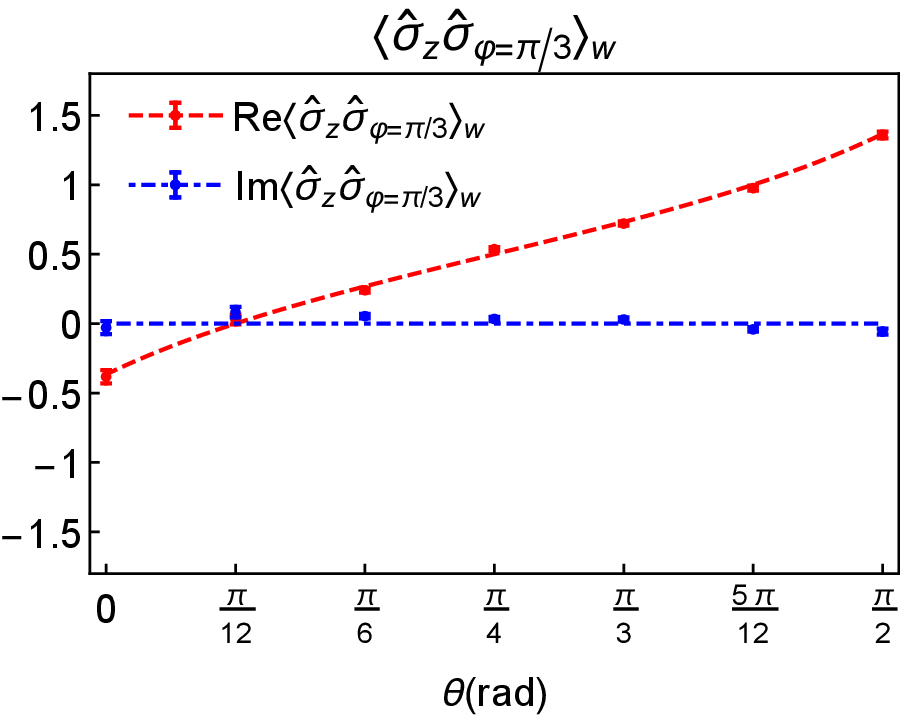}}
\subfigure[]{\label{Fig.2.subf.2}
\includegraphics[scale=0.468]{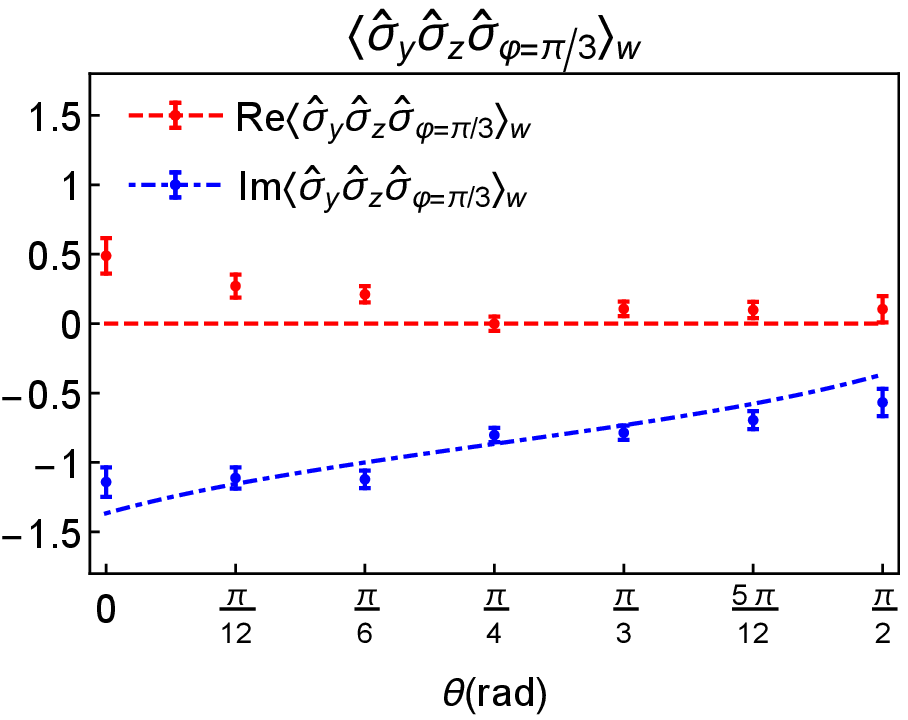}}
\caption{Sequential weak value $\langle\hat{\sigma}_{y}\hat{\sigma}_{z}\hat{\sigma}_{\varphi=\pi/3}\rangle_{w}$ with coupling parameter $\gamma=25^{\circ}$.  }
\label{Fig.2.3}

\subfigure[]{\label{Fig.2.subf.1}
\includegraphics[scale=0.468]{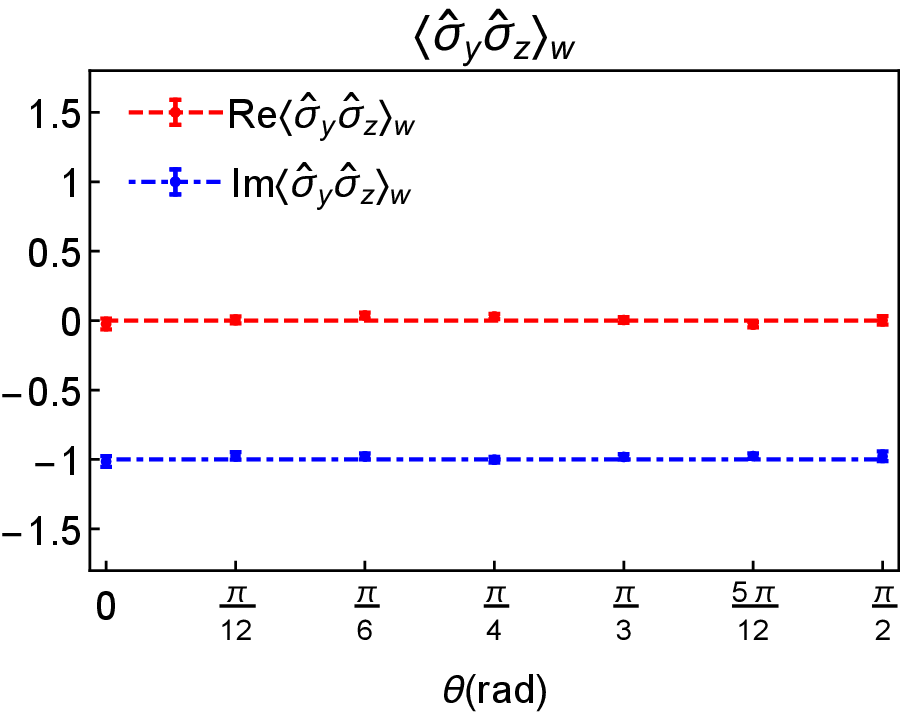}}
\subfigure[]{\label{Fig.2.subf.2}
\includegraphics[scale=0.468]{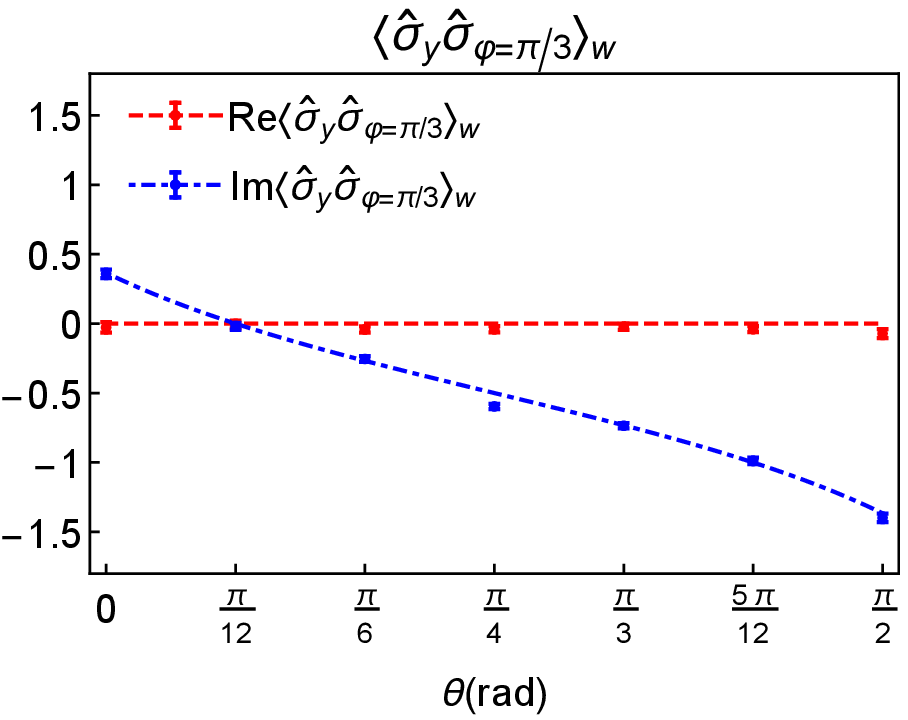}}
\subfigure[]{\label{Fig.2.subf.2}
\includegraphics[scale=0.468]{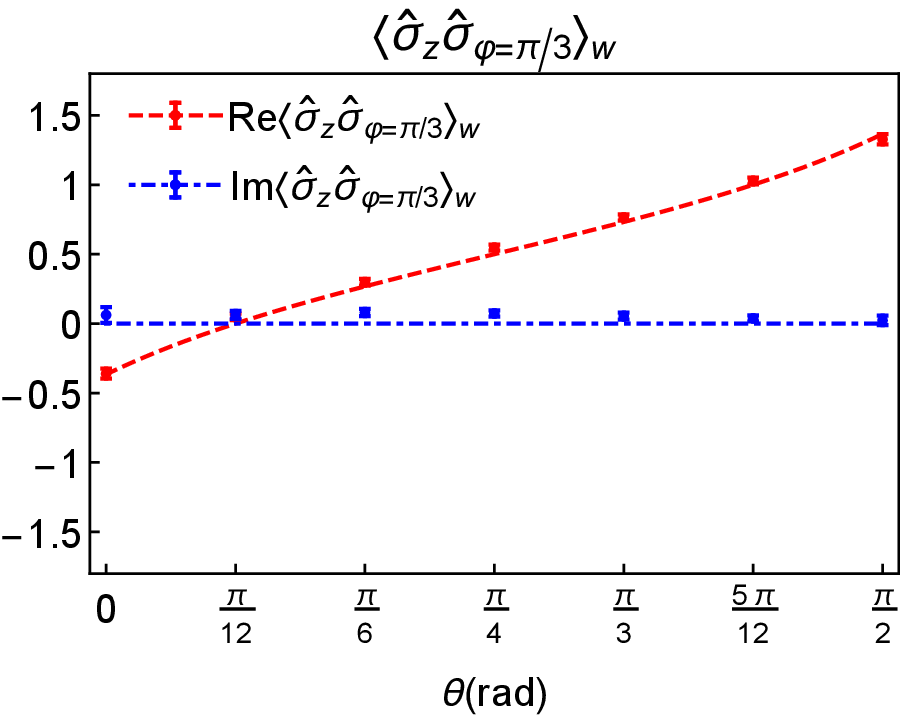}}
\subfigure[]{\label{Fig.2.subf.2}
\includegraphics[scale=0.468]{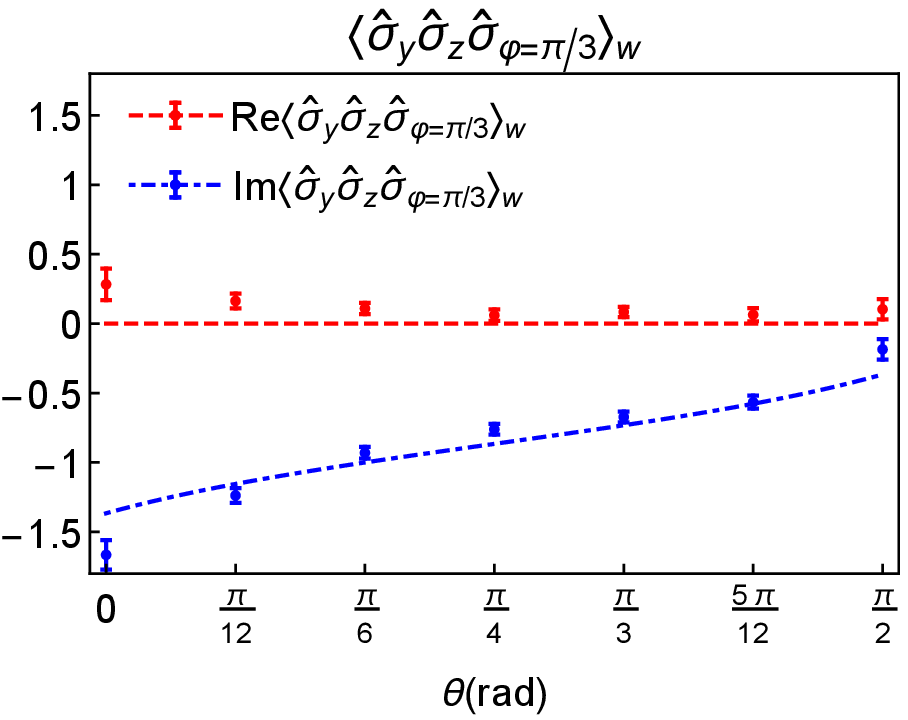}}
\caption{Sequential weak value $\langle\hat{\sigma}_{y}\hat{\sigma}_{z}\hat{\sigma}_{\varphi=\pi/3}\rangle_{w}$ with coupling parameter $\gamma=30^{\circ}$.  }
\label{Fig.2.4}
\end{figure*}

Now we consider SWMs of arbitrary $N$ observables $\hat{A}_{1},\hat{A}_{2},...,\hat{A}_{N}$. The state of the composite system, assuming that weak coupling strength $\theta$ is the same for every weak measurement without loss of generality, becomes
\begin{equation}
|\Psi\rangle_{sp_{1}...p_{N}}=e^{-i\gamma\hat{A}_{N}\otimes\hat{\sigma}_{y}}\cdot\cdot\cdot e^{-i\gamma\hat{A}_{1}\otimes\hat{\sigma}_{y}}|\psi_{i}\rangle|0_{p_{1}}\rangle\cdot\cdot\cdot|0_{p_{N}}\rangle.
\end{equation}
After SWMs, post-selecting system on $|\psi_{f}\rangle$ gives the state of pointers as
\begin{equation}
|\Phi\rangle_{p_{1}...p_{N}}=\langle\psi_{f}||\Psi\rangle_{sp_{1}...p_{N}}.
\end{equation}
Here SWV is defined as $\langle\hat{A}_{1}\cdot\cdot\cdot\hat{A}_{N}\rangle_{w}=\langle\psi_{f}|\hat{A}_{N}\cdot\cdot\cdot\hat{A}_{1}|\psi_{i}\rangle/\langle\psi_{f}|\psi_{i}\rangle$, which can be obtained by performing measurement on $N$ pointers (see details in Supplementary, Part A). In the case of $N=3$, we obtain
\begin{equation}
\begin{split}
\langle\hat{\sigma}_{+}\hat{\sigma}_{+}\hat{\sigma}_{+}\rangle_{p_{1}p_{2}p_{3}}&=2\gamma^{3}\mathrm{Re}[\langle\hat{A}_{1}\hat{A}_{2}\hat{A}_{3}\rangle_{w}+\langle\hat{A}_{1}\hat{A}_{2}\rangle_{w}\langle\hat{A}_{3}\rangle^{\dagger}_{w} \\
&+\langle\hat{A}_{1}\hat{A}_{3}\rangle_{w}\langle\hat{A}_{2}\rangle^{\dagger}_{w}+\langle\hat{A}_{2}\hat{A}_{3}\rangle_{w}\langle\hat{A}_{1}\rangle^{\dagger}_{w}]. 
\end{split}
\end{equation}
Similarly, the imaginary part of SWV is determined by
\begin{equation}
\begin{split}
\langle\hat{\sigma}_{R}\hat{\sigma}_{R}\hat{\sigma}_{R}\rangle_{p_{1}p_{2}p_{3}}&=2\gamma^{3}\mathrm{Im}[-\langle\hat{A}_{1}\hat{A}_{2}\hat{A}_{3}\rangle_{w}+\langle\hat{A}_{1}\hat{A}_{2}\rangle_{w}\langle\hat{A}_{3}\rangle^{\dagger}_{w} \\
&+\langle\hat{A}_{1}\hat{A}_{3}\rangle_{w}\langle\hat{A}_{2}\rangle^{\dagger}_{w}+\langle\hat{A}_{2}\hat{A}_{3}\rangle_{w}\langle\hat{A}_{1}\rangle^{\dagger}_{w}]. 
\end{split}
\end{equation}
It is very interesting to note that the real part and imaginary part of SWVs depend on the number of $\hat{\sigma}_{R}$ in the expectation value \cite{sequential}. The real part of SWV is obtained if there is an even number of $\hat{\sigma}_{R}$, otherwise imaginary part of SWV is obtained.

In the case that the measured observable is Pauli type i.e. $\hat{A}\equiv\hat{\sigma}_{A}=\vec{\sigma}\cdot\vec{n}_{A}$, we could obtain the weak value $\langle\hat{\sigma}_{A}\rangle_{w}$ without approximation since that $e^{-i\gamma\hat{\sigma}_{A}\otimes\hat{\sigma}_{y}}=\mathrm{cos}\gamma-i\mathrm{sin}\gamma(\hat{\sigma}_{A}\otimes\hat{\sigma}_{y})$. The respective measurement of $\hat{\sigma}_{+}, \hat{\sigma}_{R}$ on the pointer gives
\begin{equation}
\begin{split}
\langle\hat{\sigma}_{+}\rangle_{p}&=\dfrac{\mathrm{sin}(2\gamma)\mathrm{Re}(\langle\hat{\sigma}_{A}\rangle_{w})}{\mathrm{cos}^{2}(\gamma)+\mathrm{sin}^{2}(\gamma)|\langle\hat{\sigma}_{A}\rangle_{w}|^{2}}  \\
\langle\hat{\sigma}_{R}\rangle_{p}&=\dfrac{\mathrm{sin}(2\gamma)\mathrm{Im}(\langle\hat{\sigma}_{A}\rangle_{w})}{\mathrm{cos}^{2}(\gamma)+\mathrm{sin}^{2}(\gamma)|\langle\hat{\sigma}_{A}\rangle_{w}|^{2}},
\end{split}
\end{equation}
which reduces to Eq. (3) naturally in the first order approximation. The exact SWVs of multiple Pauli observables are also obtainable (see details in Supplementary, Part B). The exact expressions show that weak values and SWVs are independent of measurement strength in the case that Pauli type observables are measured.

{\it Experimental realization of sequential weak measurements of photons.}
Fig. \ref{fig.1} shows the experimental setup of realizing SWMs of polarization observables of single photons. The setup consists of five parts, i.e. heralded single photons source, initial state preparation, sequential weak measurements, post-selection and detection.

The single photons source is realized by heralding the coincidence of pair of photons. The pair of photons is  generated via spontaneous parameter down conversion (SPDC) process by pumping a 2 mm thick type-I beta barium borate (BBO) crystal with a 808 nm mode-lock Ti:Sapphire laser (repetition rate : 76 MHz).
After coupled to single-mode fibre (SMF),  the idler photons are directly sent to silicon single-photon avalanche detector (SPD) while the signal photons is connected to a launcher and then emitted along the free-space path. The signal photons are prepared into the initial state $|\psi_{i}\rangle=(|H\rangle+|V\rangle)/\sqrt{2}$ by using a polarizing beam splitter (PBS) and a half wave plate (HWP) rotated at $\pi/8$ after the PBS, where $|H\rangle$ and $|V\rangle$ represent the horizontal and vertical polarization state respectively. The sequential weak measurements of polarization observables of the signal photons is carried out via the weak measurement module a, b and c as shown in Fig. 1. After passing through the module c, the signal photons is projected into final state $|\psi_{f}\rangle=\mathrm{cos}\theta|H\rangle+\mathrm{sin}\theta|V\rangle$  via a HWP rotated at $\theta/2$ and a PBS. The signal photons, after the PBS, are then coupled to a SMF and sent to a SPD for coincidence detection.

The key of our setup is the weak measurement module that performs weak measurement of arbitrary polarization observable.  
Here we take the module c, which realizes weak measurement of observable $\hat{\sigma}_{\varphi}\equiv|\varphi\rangle\langle\varphi|-|\varphi^{\perp}\rangle\langle\varphi^{\perp}|$ with $|\varphi\rangle=\mathrm{cos}\varphi|H\rangle+\mathrm{sin}\varphi|V\rangle$ and $|\varphi^{\perp}\rangle=\mathrm{sin}\varphi|H\rangle-\mathrm{cos}\varphi|V\rangle$, as an example to explain how it works.
The basic idea is that we first transforms measurement basis $\lbrace|\varphi\rangle, |\varphi^{\perp}\rangle\rbrace$ into $\lbrace|H\rangle, |V\rangle\rbrace$ via H3 and then performs weak measurement of observable $\hat{\sigma}_{z}\equiv|H\rangle\langle H|-|V\rangle\langle V|$ by using optical elements between H3 and H23 \cite{hu}. We complete weak measurement of $\hat{\sigma}_{\varphi}$ by transforming system back to measurement basis $\lbrace|\varphi\rangle, |\varphi^{\perp}\rangle\rbrace$ via H23 at last. Both H3 and H23 are rotated at $\varphi/2$.
Suppose that photons prepared in the polarization state $\alpha|\varphi\rangle+\beta|\varphi^{\perp}\rangle$ are sent into the module c. After passing through the H3, the state of photons is transformed into $\alpha|H\rangle+\beta|V\rangle$. In order to realize weak interaction $e^{-i\gamma\hat{\sigma}_{z}\otimes\hat{\sigma}_{y}}$, we encode the information of system i.e. polarization of photons into the degree of freedom of optical path and use the degree of freedom of polarization as the pointer. This is realized by using a beam displacer (BD), the H17 rotated at $\pi/4$ and the H18 rotated at $-\gamma/2$, the H19 rotated at $\gamma/2$. The parameter $\gamma$, which reflects the strength of measurement, can be continuously adjusted in our case. The composite state of photons, after weak interaction, becomes $\alpha|0\rangle\otimes(\mathrm{cos}\gamma|H\rangle-\mathrm{sin}\gamma|V\rangle)+\beta|1\rangle\otimes(\mathrm{cos}\gamma|H\rangle+\mathrm{sin}\gamma|V\rangle)$, where the path states $|0\rangle, |1\rangle$ represent photons flying along the down arm and up arm respectively. We then perform projective measurements on the pointer via the Q5, the H20 and a PBS. The information of system that encoded in the path basis $\lbrace|0\rangle, |1\rangle\rbrace$ need to be recorded back into polarization basis for subsequent measurement, which is realized by recombining light of two arms via the H21 rotated at $\pi/4$ placed in down arm and a BD. The H22 rotated at $\pi/4$ after the BD is used for obtaining the correct state form. The system is transformed back to measurement basis $\lbrace|\varphi\rangle, |\varphi^{\perp}\rangle\rbrace$ via H23 and this finishes the weak measurement of observable $\hat{\sigma}_{\varphi}$.  

The measurement of the pointer for each weak measurement, as described above, is realized via a polarizing analyzer placed between BDs in module. The outcome of measurement is encoded in the path of outgoing photons. By implementing projective measurement of basis state of observable separately, the expectation value of the observable is calculated by combining outcomes of projective measurements. The expectation value of multiple observables, which is required to obtain SWVs, can be obtained similarly.  

{\it Results.}
In our experiment, the module a, b and c realize weak measurement of polarization observable $\hat{\sigma}_{y}$, $\hat{\sigma}_{z}$ and $\hat{\sigma}_{\varphi}$ respectively. Here $\hat{\sigma}_{y}\equiv |R\rangle\langle R|-|L\rangle\langle L|$, $\hat{\sigma}_{z}\equiv |H\rangle\langle H|-|V\rangle\langle V|$. We perform the measurement with $\hat{\sigma}_{\varphi=\pi/3}$ case such that SWV of three non-commuting observables, i.e., $\langle\hat{\sigma}_{y}\hat{\sigma}_{z}\hat{\sigma}_{\varphi=\pi/3}\rangle_{w}$ can be obtained as shown in Fig. 2 and Fig. 3. The measurement strength of all three modules, which is determined by parameter $\gamma$, is taken with the same value in our experiment. Two different cases with $\gamma=25^{\circ}$ and $\gamma=30^{\circ}$ are considered in our experiment as shown in Fig. 2 in comparison with Fig. 3 , which verify that SWVs are independent of measurement strength when Pauli type observables are measured.
Of course, each weak measurement module is allowed to has different measurement strength but it will not affect the measured sequential weak values. Since $\gamma$ can be continuously tuned via rotating HWP, the modules with $\gamma=0$ perform no measurement at all, which enable us to realize weak measurement of arbitrary observable and SWMs of arbitrary two observables. The Fig. 2 and Fig. 3 show the measured sequential weak values of arbitrary two observables and three observables in the case of different post-selected states $|\psi_{f}\rangle=\mathrm{cos}\theta|H\rangle+\mathrm{sin}\theta|V\rangle$. The weak value of arbitrary observable and SWVs of pauli observables $\hat{\sigma}_{x}, \hat{\sigma}_{y}, \hat{\sigma}_{z}$ are also measured (see Supplementary Material).  
Considering the statistical error and possible imperfections of optical elements, our results fit well with theoretical predictions. The results of three sequential measurements are worse than the case of two is due to the fact that system errors will be accumulated when more measurement modules are added.

{\it Discussion and conclusion.}
Although we have only demonstrated sequential weak measurements of three non-commuting observables, our proposal allows measurement of arbitrary observables by using suitable number of weak measurement modules. The controllable measurement strength and modularize design of weak measurement make our proposal suitable for various weak measurements task with photonic system. 

In conclusion, we have proposed how to realize sequential weak measurements of arbitrary observables and experimentally demonstrated for the first time the measurement of sequential weak values of three non-commuting observables with heralded single photons, which is impossible with the previous reported methods. The results presented here will improve further our understanding of mystery quantum world, e.g. testing quantum contextuality \cite{tex1, tex2}, macroscopic realism \cite{single5, rea1, rea2}, uncertainty relation \cite{re1, re2}, verifying measurement induced geometric phase \cite{geo1} and so on \cite{Feyn, D1}. Moreover, our results have important applications in the areas such as realization of counterfactual computation \cite{sequential, coun1, coun2}, direct measurement of matrix density \cite{7, Lundeen}, direct process tomography \cite{pro2, pro1} and unbounded randomness certification \cite{ran1, ran2}.

This work was supported by the National Key Research and Development Program of China (No.\ 2017YFA0304100, No. 2016YFA0301300 and No. 2016YFA0301700), NNSFC (Nos.\ 11774335, 11474268, 11504253, 11674306 and 61590932), the Key Research Program of Frontier Sciences, CAS (No.\ QYZDY-SSW-SLH003), the Fundamental Research Funds for the Central Universities, and Anhui Initiative in Quantum Information Technologies (Nos.\ AHY020100, AHY060300). J. S. Chen and M. J. Hu contribute equally to this work.

\end{document}